\documentclass[aps,final,reprint,twocolumn,floatfix,showpacs,superscriptaddress,notitlepage]{revtex4-1}

\usepackage{graphicx,amssymb,soul,color,amsmath,bm}
\usepackage{epstopdf,hyperref}
\usepackage{braket,dsfont,nicefrac}
\usepackage[mathcal]{euscript}
\usepackage{subcaption}
\hypersetup{colorlinks=true,citecolor=blue,linkcolor=magenta}

\sethlcolor{yellow}
\allowdisplaybreaks

\usepackage{xcolor}

\begin{document}

%\title{Origin of the low-energy optical excitations in NiPS$_3$}
\title{Singlet polaron theory of low-energy optical excitations in NiPS$_3$} %%% NO SE...

\author{I. J. Hamad}
\affiliation{Instituto de Física Rosario (CONICET) and Facultad de Ciencias Exactas, 
Ingeniería y Agrimensura, Universidad Nacional de Rosario, 2000 Rosario, Argentina}

\author{C. S. Helman}
\affiliation{Instituto de Nanociencia y Nanotecnolog\'{\i}a CNEA-CONICET,
Centro At\'{o}mico Bariloche and Instituto Balseiro, 8400 Bariloche, Argentina}

\author{L. O. Manuel}
\affiliation{Instituto de Física Rosario (CONICET) and Facultad de Ciencias Exactas,
Ingeniería y Agrimensura,  Universidad Nacional de Rosario, 2000 Rosario, Argentina}

\author{A. E. Feiguin}
\affiliation{Physics Department, Northeastern University, Boston, MA 02115, USA}

\author{A. A. Aligia}
\affiliation{Instituto de Nanociencia y Nanotecnolog\'{\i}a CNEA-CONICET,
Centro At\'{o}mico Bariloche and Instituto Balseiro, 8400 Bariloche, Argentina}

\begin{abstract}
%
%Light-matter interactions in quasi two-dimensional magnetic systems are intensively studied due
%to their potential technological applications. They present an opportunity to realize novel many-body states of matter, and to study the interplay between electronic and magnetic degrees of freedom. In particular, tightly bound many-body states that behave as coherent quasi-particles are rare, and may lead to unconventional technological applications beyond the
Light-matter interactions can be used as a tool to realize novel many-body states of matter and to study the interplay between electronic and magnetic degrees of freedom. In particular, tightly bound many-body states that behave as coherent quasi-particles are rare, and may lead to unconventional technological applications beyond the
semi-conductor paradigm, particularly if these excitations are bosonic and can condense. Two-dimensional magnetic systems present a pristine platform to realize and study such states.
We construct a theory that explains the low-energy optical excitations at 1.476 eV and
1.498 eV observed by photoluminiscence, optical absorption, and RIXS in the van der Waals
antiferromagnet NiPS$_3$. Using \textit{ab initio} methods, we construct a two-band Hubbard
model for two \textit{effective} Ni orbitals of the original lattice. The dominant effective
hopping corresponds to third-nearest neighbours. This model exhibits two triplet-singlet
excitations of energy near two times the Hund exchange. 
From perturbation theory, we obtain
an effective model for the movement of the singlets in an antiferromagnetic background, that we solve using a generalized self-consistent Born approximation. These singlet excitations, dressed by a cloud of magnons, move coherently as polaronic-like quasi-particles, ``singlet polarons”.  Our theory explains the main features of the observed spectra.

\end{abstract}

%\pacs{75.20.Hr, 71.27.+a, 71.10.-w} El uso de los PACS cayó en desuso.

\maketitle

\section{Introduction}
\label{intro}

\begin{figure*}[ht]
\begin{subfigure}{0.3\textwidth}
    \centering
    \includegraphics[scale=0.25]{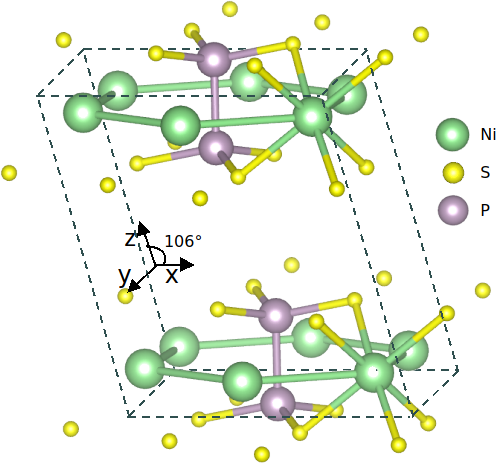}
    \caption{}
    \label{fig:struct}
\end{subfigure}
\begin{subfigure}{0.3\textwidth}
    \centering
    \includegraphics[scale=0.18]{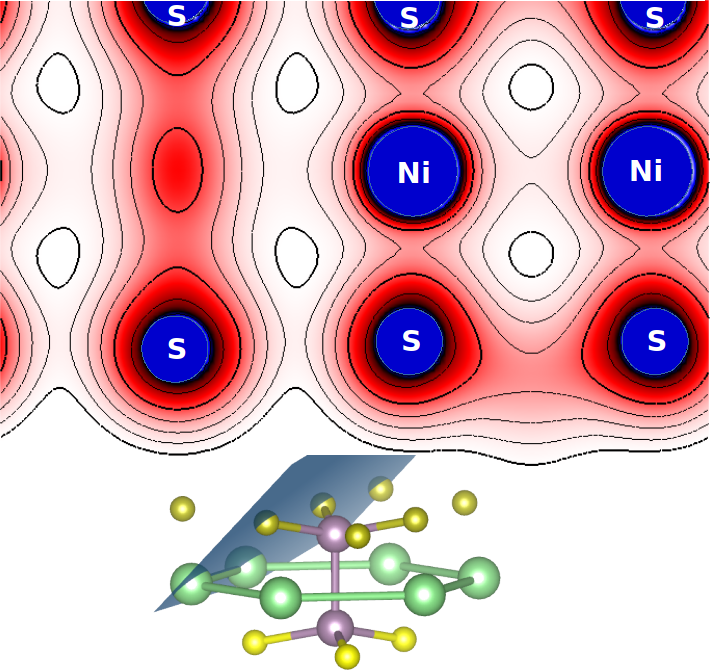}
    \caption{}
    \label{fig:dens1}
\end{subfigure}
\begin{subfigure}{0.3\textwidth}
    \centering
    \includegraphics[scale=0.17]{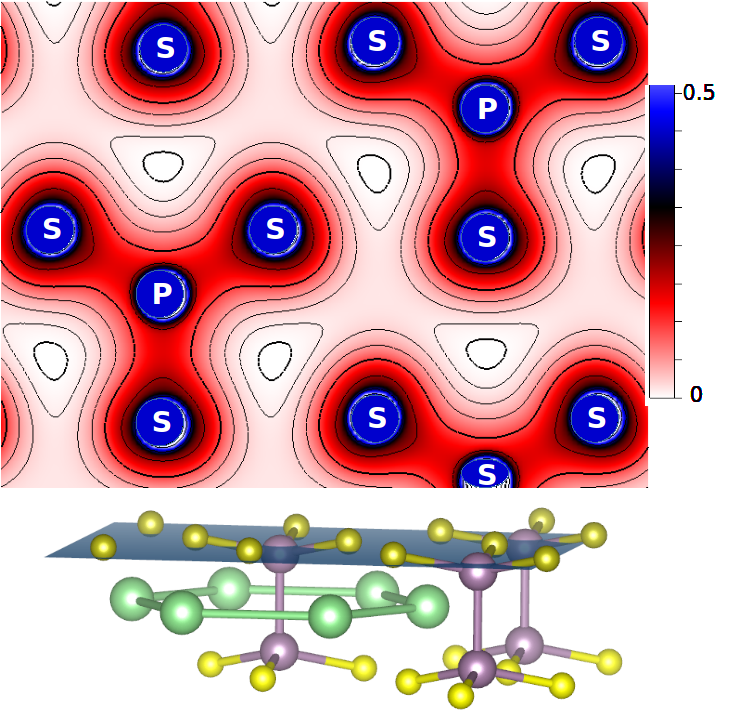}
    \caption{}
    \label{fig:dens2}
\end{subfigure}
\caption{(Colour online) (a) Structure of bulk NiPS$_3$. Note the P dimmer bonded to S atoms at the centre of the Ni hexagons. (b) Valence charge density for two different planes as sketched in each case. Note the difference in the charge density along bonds in each case. The colour scale unit is in electrons/\AA$^3$\cite{vesta}.
}
\label{fig1}
\end{figure*}
For many years, the idea of an archetypal two-dimensional (2D) magnetic material was
a dream of theorists and experimentalists alike. The physics in low dimensional
systems differs in many aspects from the three-dimensional counterpart, mostly due to
a more pervasive and dominant role of quantum fluctuations, that prevent spontaneous
symmetry breaking and thus, true long range order, from taking place in absence of
anisotropy~\cite{Mermin1966}. 
This offers the possibility of realizing exotic quantum states of matter, such as quantum
spin liquids \cite{Savary2017}. Still, different types of magnetic order are
possible in 2D depending on the interactions and anisotropies present in the
system~\cite{Gibertini2019,Lee2016,Mehlawat2022,Olsen2021,Afanasiev2021,Qi2023}. The recent discovery of atomically thin exfoliated
van der Waals magnets~\cite{Kuo2016,Lee2016,Burch2018,Gibertini2019}, particularly of the
family of MPX$_3$ metal phosphorous trichalcogenides (M=Mn, Fe, and Ni), has opened the
research field in new unexplored and exciting directions, that also have important technological implications. 

While many of these compounds are ferromagnets~\cite{Lee2016}, NiPS$_3$ was found to realize
an unconventional zigzag antiferromagnetic order, first observed in its bulk
form~\cite{Joy1992,Wildes2015}, and also in quasi-2D multi-layer exfoliated
structures~\cite{Kuo2016}, enabling studies of the dimensional crossover of the magnetic
order~\cite{Kim2019,plumley2023,DiScala2023,Kim2021,Afanasiev2021}. The minimal model used to understand the
magnetic properties of this material consists of a generalized Heisenberg model with spins
$S=1$ sitting in place of the Ni$^{2+}$ ions, at the vertices of a honeycomb
lattice~\cite{Olsen2021}. While the low-energy magnetic excitations in NiPS$_3$ seem to be
well described by spin-wave 
theory~\cite{Scheie2023}, this material has been shown to
exhibit strong interactions between spin-ordering and optical excitations~\cite{Kang2020,Kim2018,Kim2023,Wang2021,Dirnberger2022,Jana2023,Allington2024}.   
In particular, several recent optical experiments [photoluminiscence, optical absorption, 
and resonant inelastic X-ray scattering (RIXS)]~\cite{Kang2020} identified two very narrow peaks: I at 1.4756 eV and II at 1.498 eV. The width of both peaks decrease with
decreasing temperature and for peak I, it reaches 0.4 meV well below the Néel temperature of
about 155 K. Increasing temperature, the peaks disappear together with the antiferromagnetic order. Furthermore, the lowest energy peak observed in photoluminescence is accompanied by two much lower intensity satellite peaks.

Due to this unusually narrow width, it has been proposed that peak I~\cite{Kang2020}
corresponds to a condensate of Zhang-Rice singlet (ZRS) excitons~\cite{Zhang1988}. 
These singlets are low-energy states formed between a hole in the $d$ shell of a 
transition-metal ion and another in the $p$ shell of an anion (usually oxygen) 
with the same symmetry.
They were widely discussed in the context of the cuprates~\cite{Zhang1988,Feiner1996,Hamad2018,ali20}. 
Peak II has been loosely assigned to a two-magnon sideband associated with peak I.

The interest in these excitation peaks, as well as other structures below the conduction gap, has increased in the last years \cite{Hwangbo2021,Ho2021,Wang2021,Belvin2021,Afanasiev2021,Dirnberger2022,KimKim2023,DiScala2023,Jana2023,Kim2023,Allington2024,Klaproth2023}.
Calculations including correlations are limited to a NiS$_6$ cluster. They indicate that the ground state is a triplet in a mixed
configuration between $d^8$ and $d^9L$ and the excited state possibly related to 
peak I is a singlet in the $d^9L$ configuration, compatible  with a 
ZRS~\cite{Kang2020,DiScala2023,Belvin2021,Klaproth2023,KimKim2023}.
However, 
%probably due to the complexity of the system, 
a detailed description of
the singlet, as contained for example in the studies of the 
cuprates~\cite{Zhang1988,Feiner1996,Hamad2018,ali20} has not been presented.
In addition, since these studies are limited to one Ni atom, the dynamics of the excitons is not studied.
%In particular, for the ZRS, it is essential that $d$ and $p$ holes that constitute the singlet have the same symmetry ($B_{1g}$ of the point group $D_{4h}$, like $x^2-y^2$), so that the singlet gains from $d-p$ hybridization mixing with the configurations with two $d$ and two $p$ holes  \cite{Feiner1996,ali20}.

In this work, we have combined three theoretical approaches to study the electronic
structure of the system: 
(i) we construct a low-energy Hubbard model with hopping matrix elements obtained 
by \textit{ab initio} calculations using maximally localized Wannier functions (MLWFs) 
in a suitable chosen energy range and standard on-site interactions. 
We find that two Wannier functions centered at the Ni sites, but containing a mixture of orbitals of other atoms, perfectly fit the bands near the Fermi energy with suitable hoppings. An adequate choice of Wannier functions has been shown to be important in the description of the cuprates \cite{VuFe2024}. 
This model leads naturally to two triplet-singlet excitations of energy near $2J_H$ 
where $J_H \sim 0.75$ eV is the Hund coupling. 
(ii) From perturbation theory, we construct an effective Hamiltonian $H_\text{eff}$ that describes the movement of the singlets in a spin-1 antiferromagnetic background.
(iii) Finally, we solve $H_\text{eff}$ using a generalization of the self-consistent Born approximation 
(SCBA), which is the state-of-the-art technique to tackle similar but simpler problems.
Previously, the case of a single exciton moving in Sr$_2$IrO$_4$ was mapped to a hole moving in an antiferromagnetic background \cite{Kim2014}. 
However, in the present work we have a spin-1 background and two types of singlet excitations, each of which can transform into the other. Therefore, the theory is considerably richer and more involved.

Our findings indicate that the narrow peaks I and II observed in optical experiments can be attributed to unconventional singlet polaron quasi-particles. These are triplet-to-singlet excitations that become dressed by magnons as they move, causing a distortion in the zig-zag antiferromagnetic order of NiPS$_3$. Additionally, we trace the origin of the satellite peaks to strings of misaligned spins.

\section{Results}
\subsection{Atomic and magnetic structure}
\label{ato}
The NiPS$_3$ layered compound belongs to the transition-metal phosphorus trichalcogenides (TMPS$_3$) family.
In its monolayer form, it adopts a hexagonal structure with the $D_{3d}$ point group, while the bulk form displays a monoclinic structure with the $C_{2h}$ point group, resulting from a staking of monolayers displaced in the $x$-direction, as shown in Fig.~\ref{fig1}(a).
The monolayer has a thickness of 3.18 \AA\ formed by three atomic planes, where the Ni atoms are in the central plane  surrounded by two planes of S atoms above and below.
Within the planes,  the Ni atoms form a hexagonal lattice  where each one is bounded to six S atoms and the P atoms are near the centre of the hexagon  at the height of the S planes (top and bottom) forming a P dimmer.
This P dimmer is  strongly bonded to the S atoms, forming an anion complex with a pyramidal structure.
In Fig.~\ref{fig1} (b) and (c), we present valence-charge density obtained by DFT calculation, whereby in comparison, the plane conformed  by Ni and S atoms  have more localized charge than the plane conformed by S and P atoms, supporting the presence of the anion complex at the centre of the hexagonal lattice.

Since NiPS$_3$ exhibits a relatively high Néel temperature ($155K$)\cite{Wildes2015}, the antiferromagnetic phase is stable with strong magnetic interactions.
In that phase, the magnetization axis lies along the monolayer. 
The Ni atoms form ferromagnetic zigzag chains ordered antiferromagnetically between them, as presented in Fig.~\ref{wan}(c).
Thus, the magnetic structure reveals that at first and second magnetic nearest neighbors (NNs) the Ni atoms can be either ferromagnetically or antiferromagnetically aligned, while at distances corresponding to third magnetic NNs [the atoms connected by the vectors $\delta_i$ in Fig.~\ref{wan}(c)], the correlations are antiferromagnetic. 
It is likely that the origin of this long-range magnetic interactions lies in the anion complex formed at the centre of the hexagonal lattice that acts as an effective bridge among 
third NN Ni atoms. 
These findings are further supported by the results presented in the following section.

\subsection{The effective two-band Hubbard model}
\label{2bh}

\begin{figure*}[t]
%\begin{center}
\begin{subfigure}{0.5\textwidth}
    \includegraphics[scale=0.37]{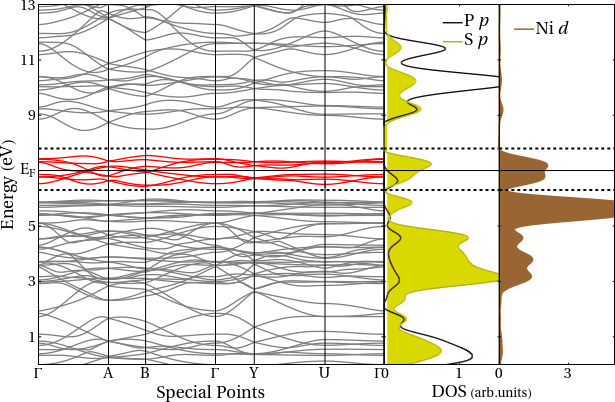}
    \caption{}
\end{subfigure}
\hfill
\begin{subfigure}{0.2\textwidth}
%   \raisebox{0.5cm}{\includegraphics[scale=0.14]{wannier1.png} }\\
%    \includegraphics[scale=0.14]{wannier2.png}
    \raisebox{0.25cm}{\includegraphics[scale=0.18]{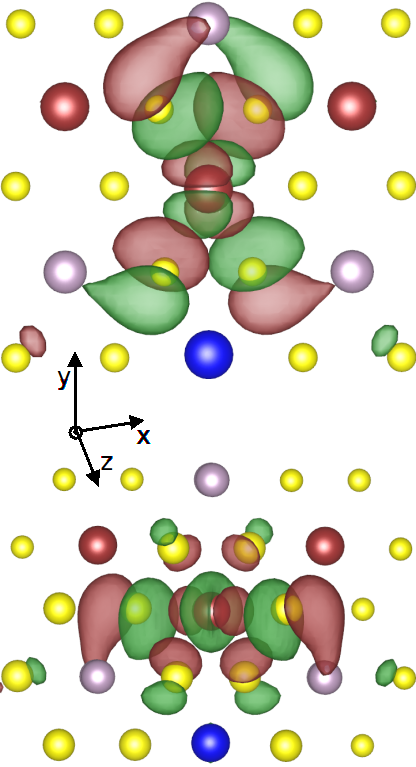}}
    \caption{}
\end{subfigure}
\hfill
\begin{subfigure}{0.28\textwidth}
   \raisebox{0.35cm}{ \includegraphics[scale=0.22]{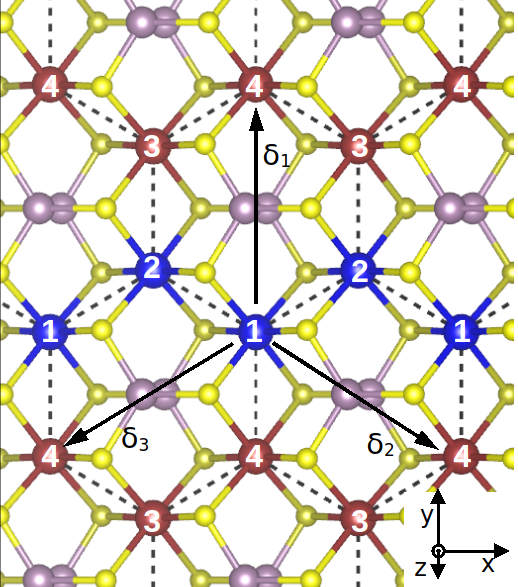} }
    \caption{}
\end{subfigure}

%\end{center}
\caption{(Colour online) (a)  Band structure of NiPS$_{3}$ for the non-magnetic case, with a unit cell that contains four unit formulae. 
In red, fit of the bands using the MLWFs. Projected density of states on different atom types are shown on the side of the band structure.
(b) Representation of the Wannier functions centred on the Ni atoms (depicted as blue/red spheres). The top/bottom panels display the odd/even Maximally Localized Wannier Functions (MLWFs), respectively. It is noteworthy that in both cases, the wave functions extend from the S atoms (yellow spheres) to the P atoms (violet spheres), indicating a strong bonding among them and displaying the mixed nature of each wave function.
%For simplicity, some atoms are not shown.
(c) Top view of the structure planes of NiPS$_3$, the numbered  red/blue spheres are the Ni atoms with spin up/down respectively. The magnetization lies in the zigzag chain direction, which is the $x$-direction. Yellow/violet spheres correspond to S and P atoms. The $\delta_i$ arrows point to the third NN 
%magnetic 
Ni atoms of Ni$_1$.}
\label{wan}
\end{figure*}
We first extract the MLWFs from a non-magnetic density-functional theory (DFT) calculation, including all relevant orbitals
in an energy window near 10 eV, obtaining 72 Wannier functions per unit cell.
The resulting hopping between Ni and S orbitals is in the range 0.5-0.9 eV, while the
corresponding hopping between P and S orbitals is much larger, between 1.5 and 2 eV. 
This value 
%one order of magnitude of difference 
indicates that P atoms play an important role and
cannot be neglected. This is also consistent with the obtained intermediate charges of the atoms, near Ni$^{+1.2}$, P$^{+1.5}$, S$^{-0.9}$.
Constructing a low-energy many-body Hamiltonian with this information is
very involved and does not allow one to single out the essential ingredients of the problem.
Therefore, we followed another strategy successfully used by two of us in another highly
covalent system~\cite{Aligia2019}. We took an energy window of 1.5 eV around the Fermi level 
containing isolated bands, as shown between dashed lines in Fig.~\ref{wan}(a).
We obtain two Wannier functions of mixed character centered at each Ni site and an 
excellent fit of the bands shown as red bands in Fig.~\ref{wan}(a). 
In the scale of the figure, no difference can be detected between the calculated bands and 
the fit. Other convergence criteria are also fulfilled as required in Ref. \cite{Pizzi_2020}.

We find that sites Ni$_1$ and Ni$_3$, which are non-equivalent in the full magnetic structure,
behave as equivalent in this reduced subspace. The same happens for Ni$_2$ and Ni$_4$.
Due to the shift of adjacent layers of the material, only one of the lattice vectors
is the axis of a $C_2$ symmetry (rotation of 180 degrees around $\delta_1$). One of the resulting MLWFs
is odd [Fig.~\ref{wan} (b) top] and the other even [Fig.~\ref{wan} (b) bottom] under
$C_2$. We denote them  by ``$o$'' and ``$e$'', respectively. 
The occupancies of both MLWFs are very close to 1 (1.03) as expected for an insulating 
system in a basis of states with small mixing with excited states. 

We denote by $\delta_1$ the vector that points from Ni$_1$ to its third NN Ni atom Ni$_4$  [see Fig.~\ref{wan} (c)] in the direction of a $C_2$ axis. 
We call $\delta_2$ and $\delta_3$ the other two vectors that connect third NNs obtained by 
rotating $\delta_1$ by (neglecting a very small distortion) 120 and 240 degrees, respectively. Surprisingly, the dominant hoppings are those between third NNs. The magnitudes are $t_1=0.2247$ eV for two $o$ orbitals at positions differing in  $\delta_1$, 
and $t_2=0.1643$ eV for $e-e$ and  $t_3=0.1048$ eV for $e-o$,  in both $\delta_2$ and 
$\delta_3$ directions. The next hopping is a NN $o-o$ one in the opposite direction of $\delta_1$, whose  value is -0.0514 eV. 
Since, as shown below, ultimately the terms governing the dynamics of the excitations
are of the order of the square of these hoppings, we start our discussion by retaining for the moment
only the third NN hoppings. In this way, the layers of the compound become divided
into four independent honeycomb sublattices.  Adding the standard interaction for two
degenerate $d$ orbitals~\cite{Aligia2004,Aligia2013}, we obtain the Hubbard model $H_{\text{Hub}}=H_{\text{hop}}+H_{\text{int}}$, with
\begin{eqnarray}
H_{\text{hop}} &=&\sum\limits_{i\in A,\sigma }[t_{1}o_{i+\delta _{1}\sigma
}^{\dagger }o_{i\sigma }+t_{2}\sum\limits_{\delta _{j}\neq \delta
_{1}}e_{i+\delta _{j}\sigma }^{\dagger }e_{i\sigma }  \notag \\
&&-t_{3}\sum\limits_{\delta _{j}\neq \delta _{1}}(-1)^{j}(e_{i+\delta
_{j}\sigma }^{\dagger }o_{i\sigma }+o_{i+\delta _{j}\sigma }^{\dagger
}e_{i\sigma })+\text{H.c.}],  \notag \\
H_{\text{int}} &=&\sum\limits_{i}[U\sum\limits_{\alpha }n_{i\alpha \uparrow
}n_{i\alpha \downarrow }
+U^{\prime }\sum\limits_{\sigma }n_{io\sigma }n_{ie\bar{\sigma}}  \notag \\
&&+(U^{\prime }-J)\sum\limits_{\sigma }n_{io\sigma }n_{ie\sigma
}-J(o_{i\uparrow }^{\dagger }e_{i\downarrow }^{\dagger }e_{i\uparrow
}o_{i\downarrow }+\text{H.c.})  \notag \\
&&+J^{\prime }(o_{i\uparrow }^{\dagger }o_{i\downarrow }^{\dagger
}e_{i\downarrow }e_{i\uparrow }+\text{H.c.})],  \label{hhub}
\end{eqnarray}
where the first sum is over one of the two sublattices (A) of the honeycomb
lattice, $\alpha $ denotes the orbitals $o$ or $e$, $n_{i\alpha \sigma
}=\alpha _{i\sigma }^{\dagger }\alpha _{i\sigma }$ and $\bar{\sigma}=-\sigma 
$.

The eigenstates and energies of $H_{\text{int}}$ at each site are easily
determined. In particular, in the most relevant subspace of two particles at
site $i$, they correspond to three singlets $|i\beta \rangle $ and one triplet 
$|itS_{z}\rangle $.  Explicitly, writing only the triplet of highest
projection $S_{z}=1$, they are 
\begin{eqnarray}
|ia\rangle  &=&\frac{1}{\sqrt{2}}\left( o_{i\uparrow }^{\dagger
}o_{i\downarrow }^{\dagger }+e_{i\uparrow }^{\dagger }e_{i\downarrow
}^{\dagger }\right) |0\rangle ,\text{ }E_{a}=U+J^{\prime },  \notag \\
|ib\rangle  &=&\frac{1}{\sqrt{2}}\left( o_{i\uparrow }^{\dagger
}o_{i\downarrow }^{\dagger }-e_{i\uparrow }^{\dagger }e_{i\downarrow
}^{\dagger }\right) |0\rangle ,\text{ }E_{b}=U-J^{\prime },  \notag \\
|ic\rangle  &=&\frac{1}{\sqrt{2}}\left( o_{i\uparrow }^{\dagger
}e_{i\downarrow }^{\dagger }-o_{i\downarrow }^{\dagger }e_{i\uparrow
}^{\dagger }\right) |0\rangle ,\text{ }E_{c}=U^{\prime }+J,  \notag \\
|it1\rangle  &=&o_{i\uparrow }^{\dagger }e_{i\uparrow }^{\dagger }|0\rangle ,
\text{ }E_{t}=U^{\prime }-J.  \label{ee}
\end{eqnarray}
For spherical symmetry (and approximately in general) $U^{\prime }=U-2J$ and 
$J^{\prime }=J.$ This leads to a triplet ground state, a singlet ($a$) at
energy $4J$ above the ground state, and two degenerate singlets ($b,c$) with
excitation energy $2J$. Since for late $3d$
$e_g$ ($t_{2g}$)
transition metal atoms a value of  $J=0.89$ (0.72) eV has been estimated, it is very natural to assume that the latter
excitations correspond to the two observed optical features near 1.5 eV.
The degeneracy can be broken by non-spherical contributions and particularly
by the displacement of the singlets in an antiferromagnetic background, as we
show below.  Note that the singlets are built from one-particle states of 
\emph{two} different symmetries, and therefore differ from conventional ZRS. For example, in cuprates the 
$d$ and $p$ orbitals that enter the ZRS both have 
$B_{1g}$ symmetry.

\subsection{Effective model for the dynamics of the triplet-singlet excitations}
\label{Heff}

Using standard perturbation theory to second-order, we construct an effective
Hamiltonian that describes the AF exchange between third NN spins $S=1$, and the movement of singlets $b,c$ in the AF background. We neglect the singlet $a$ whose 
energy is above the gap. 
In all perturbation processes, one of the two three-particle doublets of energy 
$E_{3}=U+2U^{\prime }-J$ enters the intermediate state. For simplicity, we
use bosonic operators $b_{i},c_{i},t_{im}$ ($m=S_{z}$) to represent the two
particle states of Eq. (\ref{ee}). The result is
\begin{eqnarray}
H_{\text{eff}} &=&\sum\limits_{i}\left[ (E_{b}-E_{t})b_{i}^{\dagger
}b_{i}+(E_{c}-E_{t})c_{i}^{\dagger }c_{i}\right]  + \notag \\
& + &\sum\limits_{i\in A}[J^{(a)}_{3}\mathbf{S}_{i+\delta _{1}}\cdot \mathbf{S}_{i}
+J^{(b)}_{3}\sum\limits_{\delta _{j}\neq \delta _{1}}\mathbf{S}_{i+\delta
_{j}}\cdot \mathbf{S}_{i}] + \notag \\
& + & h_{1}^{c}\sum\limits_{m}\left( c_{i+\delta _{1}}^{\dagger
}t_{im}^{\dagger }t_{i+\delta _{1}m}c_{i}+\text{H.c.}\right)  + \notag \\
&+ &h_{2}^{c}\sum\limits_{\delta _{j}\neq \delta _{1}}\sum\limits_{m}\left(
c_{i+\delta _{j}}^{\dagger }t_{im}^{\dagger }t_{i+\delta _{j}m}c_{i}
+\text{H.c.}\right)  + \notag \\
& + & h_{bc}\sum\limits_{\delta _{j}\neq \delta
_{1}}(-1)^{j}\sum\limits_{m}(c_{i+\delta _{j}}^{\dagger }t_{im}^{\dagger
}t_{i+\delta _{j}m}b_{i} + \notag  \\
&&+b_{i+\delta _{j}}^{\dagger }t_{im}^{\dagger }t_{i+\delta
_{j}m}c_{i}+\text{H.c.}),
\label{heff}
\end{eqnarray}
where
\begin{eqnarray}
J^{(a)}_{3} &=&\frac{t_{1}^{2}}{U+J},\text{ }J^{(b)}_{3}=\frac{t_{2}^{2}+2t_{3}^{2}}{U+J},
\text{ }h_{1}^{c}=\frac{t_{1}^{2}}{U-J},  \notag \\
h_{2}^{c} &=&\frac{t_{2}^{2}+t_{3}^{2}}{U-J},\text{ }h_{bc}
=\frac{t_{2}t_{3}}{2}\left( \frac{1}{U-J}+\frac{1}{U^{\prime }+J^{\prime }}\right) .
\label{param}
\end{eqnarray}
Taking  reasonable values $U=4$ eV, $J=0.75$ eV, we obtain  $J^{(a)}_{3}=10.6$
meV,  $J^{(b)}_{3}=10.3$ meV, $h_{1}^{c}=15.5$ meV, $h_{2}^{c}=11.7$  meV, and 
$h_{bc}=5.3$  meV.  The first exchange interaction excluded here is due to
NN hopping in the $\delta _{1}$ direction, and would be of magnitude 0.55
meV. At this order of magnitude, neglected states in $H_{\text{Hub}}$ 
play a role. These
exchange constant are consistent with previous studies, which indicate a
dominant role of third NN exchange of magnitude $\sim 10$ meV 
(17 meV in Ref. \onlinecite{Chittari2016} using GGA+D2+U, 9 meV in Ref. \onlinecite{Kim2018}) and larger than the NN exchange (-4 meV in Ref. \onlinecite{Chittari2016}). 
On the other hand, from a comprehensive comparison of neutron scattering experiments, DFT calculations, and linear spin wave predictions, it was obtained $J \simeq 13.9$ meV in Ref.~\onlinecite{Scheie2023}.

$H_{\text{eff}}$ already contains the main ingredients to describe the dynamics of the
triplet-singlet excitations, including the largest spin interactions and effective hoppings. 
However, in order to capture some details, like the satellites observed of the lowest lying 
photoluminescence peak~\cite{Kang2020}, one also needs to add NN exchange interactions 
$J_1$ and effective hoppings that connect the four honeycomb sublattices mentioned
previously. According to the most reliable estimates, $J_1$
$\simeq -2.7$ meV~\cite{Scheie2023}
is ferromagnetic  and (as expected) smaller in magnitude than the dominant third NN  antiferromagnetic exchange interactions $J^{(a)}_3$ and $J^{(b)}_3$. The fact that $J_1$ is ferromagnetic 
indicates the effect of a third orbital not contained in $H_{\text{Hub}}$ that contributes 
to a second-order ferromagnetic interaction through intermediate states with spin 3/2 favoured by Hund rules.

\subsection{Singlet dynamics in the zigzag order of NiPS$_3$}

To study the dynamics of the singlet excitations in the antiferromagnetic order 
of NiPS$_3$,  we have used an effective model
$H_{\text{eff}}^\prime=H_{\text{eff}}+H_{nn}$, 
where $H_{\text{eff}}$ is described above and $H_{nn}$ contains the NN interactions and hoppings (see Supplementary Material), {\it i.e.}, we work with the original structural lattice consisting of 4 honeycomb sublattices, joined by NN hoppings. 
For the latter, we have estimated that they are 10\% of the corresponding third NN values reported above. Regarding  he magnetic interactions, we have used the values obtained in Ref. \citep{Scheie2023} on the base of DFT calculations and fitting to magnetic susceptibility experiments.

In the present case, as the singlets $b$ and $c$ undergo hopping, eventually
transforming into each other, they excite two magnons within the zigzag magnetic
background. Notably, there is no direct hopping of the $b$ singlet. Due to the lack of spherical symmetry of the system, which relaxes the relation $U' = U - 2J$ as well as equal energy of odd and even Wannier functions, the degeneracy between singlets $b$ and $c$ is broken. So, we impose an energy difference $E_c - E_b$ (see Eq. \ref{heff}) between them, becoming the only completely free parameter in our calculation. This, along with dynamic effects, determines the energy positions  of the quasi-particle peaks of the singlet spectral density functions, corresponding to the observed excitonic peaks in the optical experiments~\cite{Kang2020}.

\begin{figure}[hb]
\begin {center}
\includegraphics*[width=1.0 \linewidth]{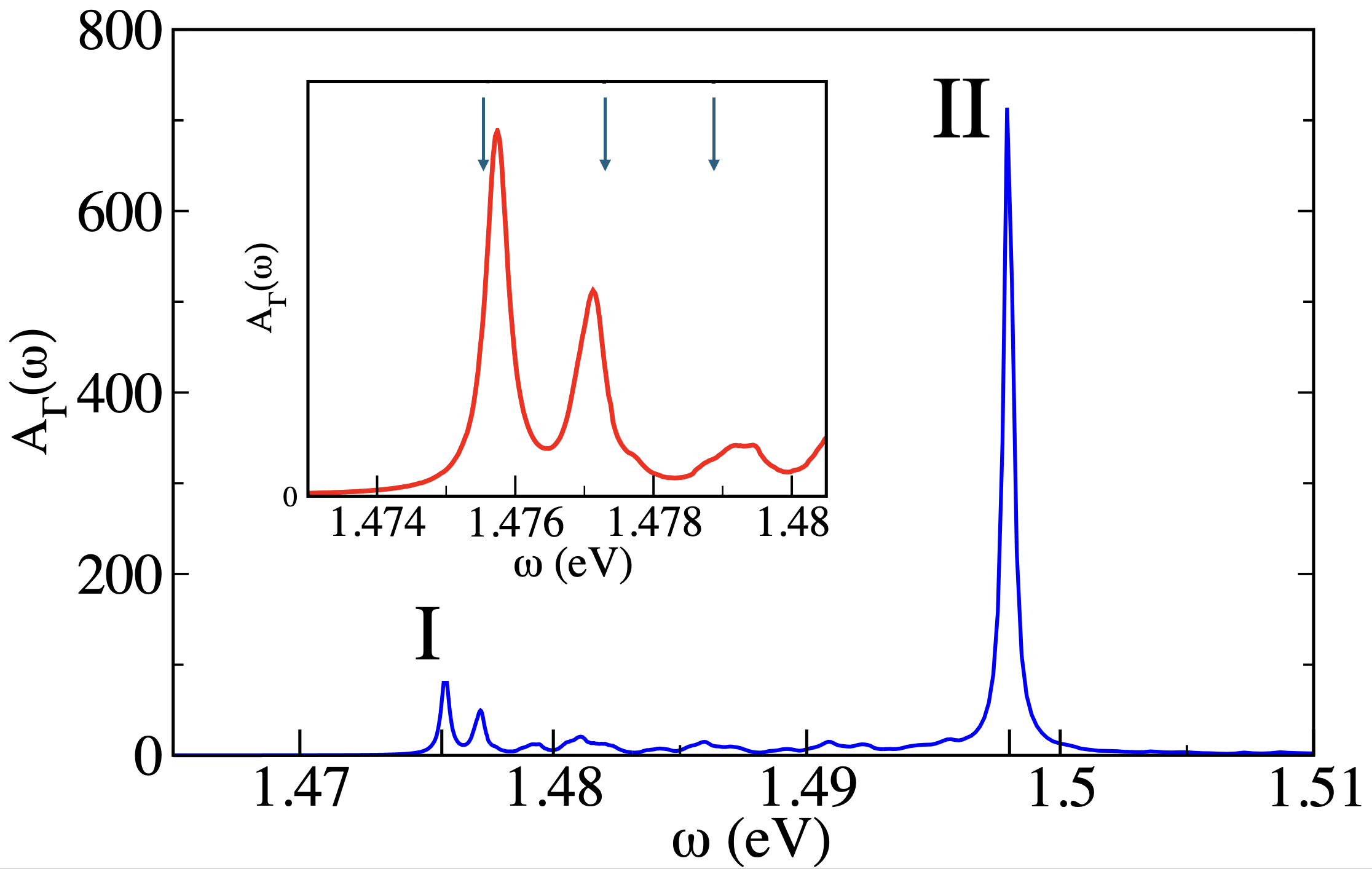}
\end{center}
\caption{(Colour online) Total spectral function of the singlets at the $\Gamma$ point as a function of energy. Two quasiparticle peaks can be observed at $1.4757$ and $1.498$ eV, together with several satellite peaks of the first. The position of the experimentally observed peaks is marked as I and II (see text for details). Inset: detail of the low energy peak and first satellites. The position of the experimentally observed main peak (peak I) and satellites is marked with arrows. }
\label{spec}
\end{figure}

We have generalized the self-consistent Born approximation to compute the
spectral function of the two relevant singlets of the effective model ($b,c$) moving in
the zigzag magnetic ground state of NiPS$_3$. 
The SCBA, a reliable diagrammatic many-body method typically used to study a single
hole moving in an antiferromagnetic background, was here adapted to a model with two ``hole'' states, each corresponding to one of the two singlets of the effective model,
exciting two magnons as they move above the antiferromagnetic order (see Supplementary Material for details of the calculation). The optimal energy difference $E_c-E_b$ that gives the results described below was found to be $E_c-E_b=29.5$ meV, quite close to the energy difference between the peaks I and II in optical experiments~\cite{Kang2020}, approximately 23 meV.

The spectral signatures of the photoluminiscence and optical absorption experiments
correspond to the $\bm k = \bm 0$ singlet spectral functions. This is because photons
have nearly zero wave vector and momentum, which enforces excitations to occur at wave
vector $\bm k = \bm 0$. 
Consequently, in Figure~\ref{spec} the calculated spectral function for $\bm k=\bm 0$ is
displayed. The results show two quasi-particle (QP) peaks at 1.4757 and 1.498 eV,
associated with singlets $c$ and $b$ respectively, and several replicas above the low 
energy peak. These QP peaks are the spectral fingerprints of singlet polaron excitations, that is, singlets coherently moving along the lattice, dressed by the magnon excitations of the zig-zag antiferromagnetic background. 

The first two replicas have an energy of 1.4771 and 1.4793 eV, 
respectively. This is in excellent agreement with the optical experiments from Ref.~\onlinecite{Kim2019}, 
that observe a peak I at an energy of 1.4756 eV, a peak II at almost 1.5 eV 
and further replicas at 1.4773 and 1.4789 eV. The origin of the the observed widths and intensities
is discussed below.

We have verified that the replicas or shoulders of the low-energy peak originate from
both the two-band structure of magnon excitations (resulting from a two-spin unit cell)
and the first NN hopping, that connect the four honeycomb sublattices.
These shoulders are the so-called ``strings'', {\it i.e.} the spectral representation of the singlets leaving behind a string of  misaligned spins while traversing the 
lattice~\cite{Liu1992}. These strings are eventually repaired by quantum fluctuations of the magnetic background. 
The distortion of the zigzag magnetic background due to the motion of the singlets, which
is much larger for the $c$ singlet, explains the fact that the intensity is considerably
lower for the low-energy singlet $c$ than for the singlet $b$. 

Since our calculation is performed at zero temperature, the singlet polaron peaks have zero linewidths,
but they are artificially  broadened by means of a small imaginary part added to the
poles of the singlet self-energies. 
In the real system, radiative decay to the ground state should take place, broadening the higher energy peak \cite{Allington2024,Andre91}, as it is observed in photoluminiscence experiments~\cite{Kim2019}.

\begin{figure}[h]
\begin {center}
\includegraphics*[width=1.0 \linewidth]{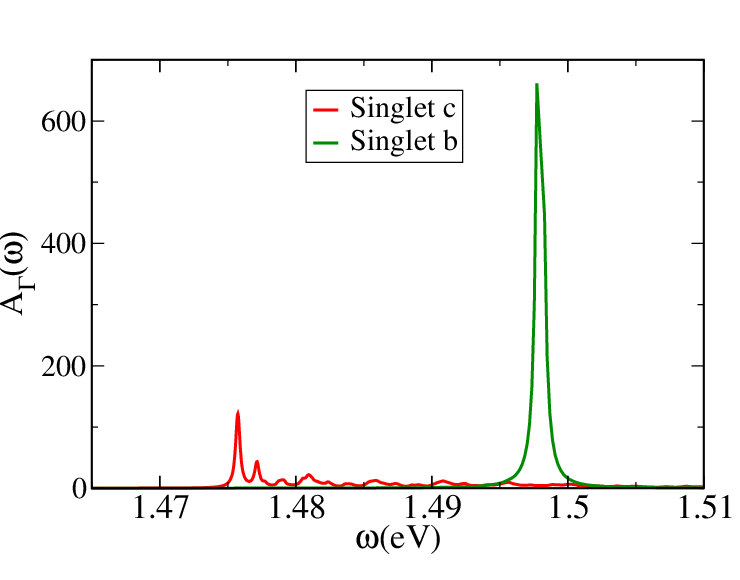}
\caption{(Colour online) Contribution to the spectral function of the singlets $b$ and $c$. 
%The spectral contributions of the $b$ singlet is that of an isolated orbital ({\it i.e.} a Dirac delta function).
}
\label{spec_b_c}
\end{center}
\end{figure}

In Figure~\ref{spec_b_c} we show the contributions of the two singlets to the spectral 
function of Fig.~\ref{spec}. We recall that singlet $c$, with larger hopping matrix elements [see below Eq. (\ref{param})] , can hop while either transforming to a $b$ singlet, or not. This translates into a quasi-particle peak accompanied with some shoulders or replicas, and an incoherent structure at higher energies. 
Moreover, the QP peak is located at an energy largely renormalized with respect to the bare
energy assigned to it (1.5296 eV). On the other hand, singlet $b$ has a spectrum that
consists of a Dirac delta-like peak centered at the bare energy of the orbital, almost 1.5 eV, indicating that this singlet does not couple to the magnetic excitations and thus behaves as a free 
 non-interacting quasiparticle. 
Consequently, it is prone to the above-mentioned radiative decay. This fact naturally explains why in the experiments it appears broader than the low energy peak. 

Since these results correspond to the spectral function of the singlet
excitations, it should be noted that, when comparing with optical experiments, 
 the relative intensities of the observed peaks are affected by matrix elements. 
The transition from a local triplet present in the ground state to the local singlet $b$ or
$c$ is made possible by spin-orbit coupling, and the orbital content is clearly different
for both singlets. This can explain the difference in intensities between our calculation and that of the experiments.  
%However, note that the relative intensities between peaks I and II have the same trend in our calculation as in the optical experiments of Ref.~\onlinecite{Kang2020}.  

\section{Discussion}

Starting from maximally localized Wannier functions near the Fermi level, we have derived an effective two-band Hubbard model that accounts for key features of the spin and charge dynamics of NiPS$_3$. The physics is dominated by a third-neighbour hopping term between ``effective'' Ni states and Hund exchange, and does not require invoking a Zhang-Rice-like construction. In the strong coupling limit, the Hamiltonian reduces to the expected Heisenberg form, and naturally explains the dominant antiferromagnetic third nearest-neighbour interactions.

We caution the reader that further Wannier functions are required to elucidate additional aspects of NiPS$_3$ physics. For instance, the nearest-neighbor ferromagnetic interaction, albeit smaller in magnitude compared to the primary  third nearest-neighbor ones estimated in our study, necessitates a third Wannier function for a spin-3/2 intermediate state.  This is likely the case also for observed d-d excitations below
the charge gap \cite{Afanasiev2021,Belvin2021,Allington2024},
and the charge gap itself.

Using simple arguments, we establish that the sharp peaks observed in several optical experiments find a natural explanation in terms of local triplet-to-singlet excitations, which engage correlated electron states within the two Ni orbitals. Despite the absence of direct spin-light coupling, these excitations are enabled by spin-orbit coupling.  The triplet-to-singlet excitations move within the Ni sublattice, distorting the spin background, giving rise to singlet polaron quasiparticles, that is, singlet states dressed by magnetic excitations.  

To solve the singlet dynamics, we apply the state of the art self-consistent Born approximation. The main features of the structure observed in optical measurements -consisting of two very narrow peaks, with the low energy one having shoulders– is reproduced in our calculation, demonstrating that the origin of these peaks is due to the formation of singlet polarons as the two singlets move on the zigzag antiferromagnetic background of NiPS$_3$. We also explain  the origin of the observed satellite peaks as strings of misaligned spins left behind by the singlet polarons as they propagate coherently on the antiferromagnetic order. 
Clearly, in our theory, the presence of the long-range magnetic order is essential to obtain the low-energy peak and their replicas. There is an evident connection between these ultrasharp resonances and the existence of antiferromagnetic order, as it was observed in the optical experiments~\cite{Kang2020}.
Further confirmation of this picture could be provided by pump-probe time-resolved ARPES experiments, as illustrated in Ref.~\onlinecite{Rincon2018}.

Although our approximation  is unable to deal with a finite concentration of singlet polarons, it is reasonable to anticipate an attractive force between them. Placing two singlet polarons in proximity, (especially the low-energy $c$ ones), results in an energy gain due to the diminished distortion of the background. This phenomenon mirrors the magnetic glue effect hypothesized in $t-J$-like models of superconductivit. Consequently, the prospect of a Bose-Einstein condensation of singlet polarons emerges as feasible.

Recent experiments have shown that the lower-energy exciton at 1.476 eV is very sensitive to the introduction of impurities \cite{Kim2023}. This might be expected since this excitation is accompanied by a cloud of spin excitations that involves several unit cells, and it should be severely affected as the average distance between impurities is smaller than the radius of the cloud. 

In conclusion, we offer a semi-quantitative explanation for the key characteristics of the remarkably narrow excitations observed below the gap in various optical experiments. These are attributed to unconventional singlet polarons: triplet-singlet excitations moving in a magnetic background. 
We calculate the dynamics of the singlet polarons 
for the first time through an advanced adaptation of the state-of-the-art SCBA method. This allows us to identify
the nature of the excited quasiparticles, which is
usually a central issue in the understanding of strongly correlated systems. Our novel insights into the interplay between magnetic order and optical excitations pave the way for deeper investigations in this realm.

\section{Methods}
The DFT calculations are performed using the \texttt{Quantum Espresso} package\cite{qe}, using a PAW pseudopotential within the GGA approximation as implemented by Perdew {\it et al.}\cite{pbe}. 
The  energy cut of the plane wave is set to 75 eV, and we use a mesh of $10\times8\times6$ points in reciprocal space. 
The NiPS$_3$ unit cell for calculations in the antiferromangetic state consist of 4 formula units with lattice parameters $a=5.819$ \AA, $b=6.621$ \AA\ and $c=10.084$ \AA.
For the Wannierization process, we use the \texttt{Wannier90} code\cite{wannier90}, selecting an energy region between 6.3 and 7.8 eV as marked with dashed lines in Fig. \ref{wan}(a).
The obtained Wannier functions are centered at Ni positions, with an excellent fit of the bands, as shown in red colour in Fig. \ref{wan}(a). 

The effective Hamiltonian is constructed using standard
methods of perturbation theory involving degenerate states.

In order to elucidate the singlet dynamics on the zig-zag antiferromagnetic order of NiPS$_3$, we have applied the diagrammatic self-consistent Born approximation (SCBA) to the above mentioned effective Hamiltonian. The SCBA calculations are explained in detail in the Supplemental Material.

\vspace{0.65cm}
\section*{Declarations}

\begin{itemize}
\item Funding: AAA is supported by PICT 2018-01546 and PICT 2020A-03661 of the Agencia I+D+i, Argentina.  LOM is supported by CONICET under grant no. 3220 (PIP2021). IJH is supported by CONICET under grant no. 0883 (PIP2021). 
CSH is supported by PICT-2021-00325 of the Agencia I+D+i, Argentina.
AEF is supported by the U.S. Department of Energy, Office of Science, Basic Energy Sciences under Award
476 No. DE-SC0022216
\item Competing interests: The authors declare no competing interests

\item Authors' contributions: C. S. Helman performed ab-initio calculations to obtain the Wannier orbitals and tight-binding parameters. A. A. Aligia coordinated the project, derived the many-body Hamiltonian and the effective model using perturbation theory. I. J. Hamad and L. O. Manuel carried out the SCBA calculations and wrote the supplementary material. A. E. Feiguin conceived the project and conducted calculations not included in the final version. All authors contributed equally to the analysis of the results and the preparation of the manuscript.

\item Acknowledgements: AEF thanks Alberto de la Torre and Kemp Plumb for illuminating discussions.
\end{itemize}

%\bibliography{references}

\end{document}